\documentstyle[11pt]{article}
\begin{document}
\begin{flushright}
\renewcommand{\textfraction}{0}
April 22th 1994\\
hep-th/9404139\\
\end{flushright}
\begin{center}
{\LARGE {\bf Models on Event Symmetric Space-Time } }
\end{center}
\begin{center}

{\Large Phil Gibbs} \\
\end{center}

%% The author welcomes all comments and corrections
%% by e-mail to phil@galilee.eurocontrol.fr

\begin{abstract}
In the event symmetric approach to quantum gravity it is assumed that the
fundamental laws of physics must be invariant under exchange of any two
space-time events. The fact that this symmetry if obviously not observed is
attributed to the possibility of a spontaneous symmetry breaking mechanism in
which the residual symmetry is diffeomorphism invariance on a continuum. The
approach has a discrete nature while dimensionality, continuity of space-time
and causality are abandoned as fundamental principles. Consequently
space-time is a dynamical object which may undergo changes of topology or
dimension. A number of different types of Event Symmetric model
are described. The most interesting models are event symmetric superstring
models based on discrete string supergroups.

\end{abstract}

\section{Introduction}

\subsection{String Theories}

Despite the lack of experimental data at ultra high scales, the search
for unified theories of particle physics
beyond the standard model has yielded many mathematical results based
purely on constraints of high symmetry, renormalisability and cancellation of
anomalies. In particular, space-time supersymmetry \cite{Wess,Zumino(1974)} has
been found to improve perturbative behaviour and to bring the gravitational
force into
particle physics. One line of research which appears to hold promise is
superstring
theory \cite{Green,Schwarz(1981a,b,1982)}. Renormalisable string models were
originally constructed in perturbative form
but were found to be incomplete in the sense that the perturbative series
were not Borel summable \cite{Gross,Perival(1988)}.

More recently there have been various attempts to formulate Superstring
Field Theories non-perturbatively (see \cite{Kaku(1991)}). The use of the
Universal
String Group to represent the symmetries and formulate a covariant action
is an elegant approach but no non-perturbative formulation has yet been
found from which the true vacuum state can be derived.

One line of investigation is to consider what might happen in string theories
when the Planck scale \cite{Planck(1899)} is reached. It is possible that
there is a phase
transition in string theories at their Hagedorn temperature
\cite{Hagedorn(1968)} near the Planck
scale. It has been speculated that above this temperature there are fewer
degrees of freedom and a restoration of a much larger symmetry
\cite{Atick,Witten(1988);Gross,Mende(1987,1988);Gross(1988)}.

One interpretation of the present state of string theories is that it
lacks a geometric foundation and that this is an obstacle to finding
its most natural formulation. It is possible that our concept of
space-time will have to be generalised to some form of stringy space.
Perhaps such space-time must be dynamical and capable of undergoing
topological or dimensional changes.

\subsection{The Loop Representation}

Recently there has been some progress in attempts to quantise Einstein
Gravity \cite{Einstein(1915a,b)} by canonical methods. A reformulation of
the classical theory in
which the connection takes the primary role instead of the
metric \cite{Ashtekar(1986,1987)} has led
to the Loop Representation of Quantum Gravity \cite{Rovelli,Smolin(1988,1990)}.

The similarity between the Loop Representation and formulations of String
Field Theories may be more than superficial
\cite{Baez(1993)}. Superstring theory and the Loop Representation can not
be equivalent since the former only works in 10 or 11 dimensions while the
latter only works in 4. It is possible that they could be different phases of
the same pre-theory provided that pre-theory allows changes of dimension.

One other remarkable aspect of the Loop Representation is that it has a
discrete
nature at scales smaller than the Planck length. This has inspired renewed
interest in discrete methods, but discrete can mean many things such as
lattice theories (see below),
spin networks \cite{Penrose(1971)} quantum set topologies
\cite{Isham,Kubyshin,Renteln(1990)}, causal sets \cite{Sorkin(1991)}
or noncommutative geometry \cite{Balachandran et al(1994)}.

\subsection{Lattice Models of Gravity}

For QCD, similar problems with the non-perturbation effects can be
formulated and analysed using Lattice Gauge Theories. The Wilson
plaquette loop action for QCD \cite{Wilson(1974)} on a lattice formulated
in the Euclidean sector
has a certain elegance since it preserves a discrete version of gauge
invariance and uses a group representation rather than a representation of a
Lie algebra. Translation invariance is also preserved in a discrete form
but the rotation group is only coarsely represented.

It is natural to ask whether a useful formulation of string theories
can be found on a lattice which could likewise improve understanding of their
non-perturbative aspects. The immediate objection to this is that a
lattice theory cannot reflect the important symmetries of string
theories in a discrete form. In particular it is not obvious how
diffeomorphism invariance could be represented. However this has been
attempted, e.g. \cite{Simon(1994)}.

Lattice studies of pure gravity start from the Regge Calculus
\cite{Regge(1961)}.
One way to retain a form of diffeomorphism invariance is to use a random
lattice on space-time instead of a regular lattice.
A random lattice does not prefer any direction.
An interesting aspect of the use of random lattices for quantum
gravity is that if the lattice is allowed to change dynamically in
some probabilistic fashion then its fluctuations merge with the
quantum mechanics in the Euclidean sector. For review see
\cite{Itzykson,Drouffe(1989)}.

There have been some encouraging results
with attempts to formulate Euclidean Gravity using dynamical triangulations.
\cite{Agishtein,Migdal(1992)}. They find evidence from numerical simulations
that 4-dimensions is critical in Euclidean gravity.
A Euclidean theory would present some
interpretation problems since it is not possible to simply apply a
Wick Rotation to a gravity theory given the dynamical aspects of the metric.

There are also encouraging results in 2+1 dimensions where Topological
Quantum Field Theories have been successfully formulated on triangulated
space \cite{Ponzano,Regge(1968),Witten(1988)}. It is found that the result
can be independent of the triangulation
so the theory must be diffeomorphism invariant. This shows that physics
can have discrete and continuous aspects at the same time. These results
are being extended to 3+1 dimensions \cite{Ooguri(1992)}. See also
\cite{Belyaev(1994)}

Whichever approach to quantum gravity is taken the conclusion is that
the Planck length is a minimum size beyond which the Heisenberg Uncertainty
Principle prevents measurement \cite{Garay(1994)}. Space-time may have
to be viewed very differently to understand physics beyond the Planck scale.
Until the geometric principles have been understood a consistent
formulation of quantum gravity may be impossible. The increase of
mathematical sophistication in physics which has emerged from these lines
of research may suggest to some that the required mathematics to solve the
problem has not yet been developed. However, the view taken here is that
a discrete approach using relatively simple mathematics may provide a
useful formulation.

\subsection{Event Symmetric Discrete Models}

The Event Symmetric approach is also discrete. It solves the problem of
diffeomorphism invariance in a different way to the random lattice approach.
The exact nature of space-time in this scheme will only become apparent
in the solution. Even the number of space-time dimensions is not set by
the formulation and must by a dynamic result.
It is possible that space-time will preserve a discrete
nature at very small length scales. The objective is to find a statistical
system which reproduces observed and hypothesised symmetries in physics and
then worry about states, observables and causality later.

We seek to formulate a lattice theory in which diffeomorphism invariance
takes a natural and explicit discrete form. At first glance it would seem that
only translational invariance can be
adequately represented in a discrete form but this overlooks the most natural
generalisation of diffeomorphism invariance in a discrete system.
Diffeomorphism invariance requires that the action should be symmetric under
any differentiable transformation of coordinates. On a discrete space we
could demand that the action is symmetric under any permutation
of the discrete space-time events ignoring continuity altogether. Generally we
will use the term {\it Event Symmetric} whenever an action has a symmetry
given by the Symmetric Group
over an infinite number of discrete "events" (or some larger group of which
it is a sub-group). In practice it will be necessary to regularise to a
finite number of events $N$ with an $S(N)$ symmetry and take the large $N$
limit.

Event symmetry is larger than the diffeomorphism invariance
of continuum space-time. If a continuum is to be restored there must be a
spontaneous mechanism of symmetry breaking in which event symmetry is replaced
by a residual diffeomorphism invariance. It stretches the imagination
to believe that a
simple event symmetric model could be responsible for the creation of
continuum space-time and the complexity of quantum gravity through symmetry
breaking, however, nature has provided an example of a similar mechanism
which may help us accept the plausibility of this claim.

Consider the way in which soap bubbles arise from a statistical physics
model of molecular forces. The forces are functions of the relative
positions and orientations of the soap and water molecules. The energy is
a function symmetric in the exchange of any two molecules of the same
kind. The system is consistent
with the definition of event symmetry since it is invariant under exchange
of any two water or soap molecules and therefore has an $S(N) \otimes S(M)$
symmetry
where $N$ and $M$ are the number of water and soap molecules. Under the right
conditions the symmetry breaks spontaneously to leave a diffeomorphism
invariance on a two dimensional manifold in which area of the bubble surface
is minimised.

A number of Event Symmetric models will be described. It is easier to
understand these as statistical theories rather than quantum theories and
it is in this form that the models are presented.
We take the optimistic view that the only difference between a statistical
event symmetric model and a quantum one is a factor of $i$ against the action
in the exponential. We hope that in the statistical version the Event Symmetry
will break to give space-time with a Euclidean signature metric while in the
quantum version it breaks to give the physical Lorentzian theory.

\section{Primitive Event Symmetric Models}

\subsection{Event Symmetric Ising Model}

The simplest event symmetric model is the event symmetric Ising model. This
consists of a large number $N$ of feromagnets represented by spin variables
\begin{equation}
                    s_i  = +1 or -1   for   (i = 1,...,N)

\end{equation}
and an action,
\begin{equation}

                    S = (\beta/N) \sum_{i<j} s_i s_j
\end{equation}
Notice the renormalisation of $\beta$ as a function of $N$.
This has $S(N)$ invariance since it is symmetric in spin permutations
and an additional $Z_2$ invariance under global spin reversal.
Solving this model is not difficult. The partition function is
\begin{equation}

                    Z = \sum_{s_i} e^{-S}
\end{equation}
Write this as a sum over states with $K$ negative spins
and $N-K$ positive spins.
\begin{equation}

             Z = {\large\sum} C(N,K) exp{\large(}(\beta/N)[(N/2)(N-1)
		    - 2 K(N-K)]{\large)}
\end{equation}

In the large $N$ limit this can be written as an integral over a variable
\begin{equation}

                         p = K/N\\

           Z = {\large\int}_0^1 dp exp{\large(}N ( \beta[1/2 - 2 p(1-p)] -
	     (p ln(p)) - (1-p)ln(1-p) ){\large)}

\end{equation}

The function in the exponential has one minimum at $p = 1/2$ for $\beta \leq 1$
and two minima for $\beta > 1$. The large $N$ limit forces the system into
these minima so there is a second order phase transition
at $\beta = 1$ with the $Z_2$ spin symmetry broken above. The $S(N)$ event
symmetry is not broken in this model.

\subsection{Event Symmetric Gauged Ising Model}

For the gauged version the spins are placed on event links. There are
therefore $(1/2)N(N-1)$ spins
\begin{equation}

                     s_{ij}  =  +1 or -1

\end{equation}

And the action is now a sum over triangles formed from three links

\begin{equation}

                    S = (\beta/N) \sum_{i<j<k}    s_{ij} s_{jk} s_{ki}
\end{equation}

This model again has an $S(N)$ event symmetry but the $Z_2$ symmetry is
now a gauge symmetry. This is already too complicated to solve exactly
by any obvious means.

The most interesting thing that can be said about this model is that it
is dual to a simple string model. By applying the usual duality
transformation on Ising models the dual action can be derived,
\begin{equation}

                    S' = ln(tanh[\beta/N]) A
\end{equation}

where the states are now closed "surfaces" made from triangles on the
lattice joined at edges. the quantity $A$ is the "Area" of the surface
defined as the number of triangles from which it is formed. This is
analogous to the Area action for a first quantised bosonic string but
is defined on an event symmetric lattice instead of a $D$ dimensional
continuous space.

There are other simple models some of which are soluble, e.g. the
spherical model.

\section{Symmetric Random Graph Models}

Since we are looking for some kind of change of dimension
it makes sense to investigate systems on which we can attempt to define
dimensionality. The simplest structure would be a random graph in
which $N$ nodes are randomly pairwise connected by links. An event
symmetric action is a function of the connections which is
invariant under any permutation of nodes. For example, actions defined
as functions of the total number of links and the total number of
triangles in a graph would be event symmetric.

The idea is that on a graph we can define dimensionality from its
connectivity. For a given node we can define a function $L(s)$, the
number of nodes which can be reached by taking at most $s$ steps
along links. If $L(s)$ has a power law on an infinite graph,
\begin{equation}

                    L(s) \rightarrow s^D as  s \rightarrow \infty

\end{equation}
then the graph has dimension D. It may also be possible to determine
dimensionality from topology of a finite graph \cite{Evako(1994)}.

If a suitable mechanism of symmetry breaking is effected on the system
the graphs generated statistically from the action may have some
finite dimension. The number of dimensions could differ from one
phase of the system to another. There could also be phases in which
the event symmetry is unbroken and the number of dimensions can be
considered infinite.

Since there are no other symmetries to guide our choice of action we
might consider heuristic criteria to contrive an action which might
exhibit spontaneous symmetry breaking of the event symmetry. As a first guess
it might be reasonable to consider an action which favours triangles
but disfavours links. The action can be written in terms of link
variables  $l_{ij}$
\begin{equation}
            l_{ij} = 1 if the nodes i and j are linked, = 0 otherwise
\end{equation}
Define
\begin{equation}
                       V_i = \sum_j l_{ij}\\

                       T_i = \sum_{j,k} l_{ij} l_{jk} l_{ki}\\

             S =  \sum_i [(\beta/N^2) T_i  -  (\alpha/N^2) V_i^2 ]
\end{equation}

A simple mean field analysis can be performed where each link is connected
with a probability $p$. Then
\begin{equation}

               T_i  ~= N^2 p^3\\

	       V_i  ~= N p
\end{equation}

Taking into account that the number density of states as a function
of p this gives an effective action of,
\begin{equation}

           S =  N[ -p ln(p) - (1-p) ln(1-p) + \beta p^3 - \alpha p^2 ]
\end{equation}

This suggests a phase transition along approximately $\beta/\alpha = 1$
with p close to one for $\beta > \alpha$ and p close to zero for
$\beta < \alpha$

Further mean field analysis of this model and other similar models is possible.
An extension to the treatment given here would be to consider a mean field
analysis of the situation where the graph breaks down into small isolated
parts. Linkage between nodes within each part can be given a probability
$p$ while linkage between nodes in different parts can be given a
probability $q$. A mean field analysis for a particular Event Symmetric
action might suggest that an asymmetric phase existed with $q$ small and $p$
close to one. It is possible that this could be taken as a signal that
other forms of Symmetry Breaking were a possibility for that action

Numerical simulations could also be used to look for evidence of
Event Symmetry breaking. It may be possible to construct models
in this way which have residual structures with finite dimensional
symmetries.

\section{Matrix Models}

If the ultimate aim is to produce event symmetric models of real physics
then it
will be necessary to introduce further symmetries such as gauge
symmetry. The Event symmetric Ising gauge model can be combined with a
random graph model giving a model with link variables which can take
three values -1, 0 or +1. Such models are interesting to study for
event symmetry breaking because the duality transformation can still
be applied to give a dual model of strings on a random graph.

To go further the $Z_2$ gauge symmetry can be extended to gauge symmetry
of other groups such as $U(1)$, $SU(3)$ etc. The link variables then takes
values zero or an element of the group. It is more natural to extend the
$Z_2$ gauge model on a random graph by allowing the link variable to take
any real value. The link variables $A_{ij}$  can be organised into the upper
triangle of a matrix. If there are no self links the diagonal terms are
zero so we extend the matrix to the lower half by making the matrix
anti-symmetric. If there are self interactions then the matrix should
be taken to be symmetric.

A four link loop action is
\begin{equation}

               S = \sum_{i,j,k,l} A_{ij} A_{jk} A_{kl} A_{lm}
\end{equation}

For antisymmetric matrices this is equal to
\begin{equation}

               S = Tr(A^4)
\end{equation}

which is an invariant under $O(N)$ similarity transformations on the matrix.

This suggests that we consider actions which are functions of the traces
of powers of the matrix $A$. Then the symmetry group of the system is $O(N)$
which has $S(N)$ as a subgroup. The idea can be extended to unitary groups
by using complex variables for hermitian matrices or symplectic groups
by using quarternions.

This is an appealing idea since it naturally unifies the $S(N)$ symmetry,
which we regard as an extension of diffeomorphism invariance, with gauge
symmetries. If the symmetry broke in some miraculous fashion then it
is conceivable that the residual symmetry could describe quantised gauge
fields on a quantised geometry.

This type of random matrix model has been extensively studied (see
\cite{Itzykson,Drouffe(1989);Kaku(1991)} )
and does not appear to produce such symmetry breaking. There are several
possible generalisations to multi-matrix models, tensor models and models
with fermions. In each case the action can be a function of any set of
scalars derived from the tensors by contraction. One interesting result
is that the $SU(N)$ matrix model in the large $N$ limit is related to a
string theory \cite{tHooft(1974)}. It is also be possible to induce
QCD from a matrix pre-theory \cite{Kazakov,Migdal(1992)}.

For the symmetry to break in the way we desire, i.e.
leaving a finite dimensional topology, the events
will have to organise themselves into some arrangement where there is an
approximate concept of distance between them perhaps defined by correlations
between field variables. Tensor variables which link events which are
separated by large distances would have to be correspondingly small. Only
variables which are localised with respect to the distance could have
significant values.

For the action to reduce to an effective local action on this space the
original action must be restricted to forms in which it is the sum of
terms which are written as contractions over tensors and which do not
separate into products of two or more such scalar quantities. For example
if there are two matrices A and B defining the field variables then
the action could contain terms such as,
\begin{equation}

                        tr(ABAB)

\end{equation}
but not,
\begin{equation}

                       tr(AB)^2
\end{equation}
or
\begin{equation}
                       tr(A)tr(B)
\end{equation}

This locality condition is important when selecting suitable actions for
models which might exhibit dimensional symmetry breaking.

It seems likely that this type of model cannot break event symmetry in any
useful way given the invariance and locality conditions described if there
are a finite number of tensors involved.

\section{Event Supersymmetric Models}

It would be an obvious next step to generalise to supersymmetric models.
So far we have matrix models based a families of groups such as $S(N)$, $O(N)$,
$SU(N)$ or $Sp(N)$. Tensor representations and invariants can be used to
construct models with commuting variables, anticommuting variables or
both. Similarly we can define models based on supersymmetry groups of
which there are also several families such as $SU(N/M)$. For analysis
of supergroups see \cite{Cornwell(1989)}

Just one simple model will
be described. The representation has an anti-hermitian matrix
$A$ of commuting variables
\begin{equation}
                   A^*_{\mu\nu} = -A_{\nu\mu}
\end{equation}
and a vector $\psi$ of anti-commuting variables.
A suitable action could be,
\begin{equation}
 S = m(2 i \psi^*_\mu \psi_\mu + A_{\mu\nu} A_{\nu\mu})\\

     + \beta( 3 \psi^*_\mu A_{\mu\nu} \psi_\nu -
        i A_{\mu\nu}A_{\nu\lambda}A_{\lambda\mu} )

\end{equation}
As well as $U(N)$ invariance this is invariant under a super-symmetry
transform with an infinitesimal anticommuting parameter $\epsilon_\nu$,
\begin{equation}
 \delta A_{\mu\nu} = \epsilon^*_\nu \psi_\mu - \epsilon_\mu \psi^*_\nu\\

	  \delta \psi_\mu = i \epsilon_\nu A_{\mu\nu}\\

	  \delta \psi^*_\mu = i \epsilon^*_\nu A_{\nu\mu}\\
\end{equation}

It is encouraging that supersymmetric generalisations of matrix models
can be so easily
constructed on event symmetric space-time. Demanding supersymmetry helps
reduce our choice of actions
but not actually very much. There are still many different possibilities
like the above which can be constructed from contractions over tensor
representations of supersymmetry groups. These models are special cases of
matrix or tensor models so they will not be more successful
as a scheme for dimensional symmetry breaking.

\section{Event Symmetric Spinor Models}

If it is not possible to break event symmetry with simple tensor models
then it is necessary to investigate models with spinor representations
or models with tensors of unlimited rank.

The advantage of spinors is that the dimension of the representations
increases exponentially with $N$. For a model using a finite number of
tensor representations the dimension is only polynomial in $N$.

A simple model would have an $O(N)$ symmetry and a Dirac spinor $\Psi$
representation with $2^{N/2}$ anticommuting components. An invariant
action can be constructed using the gamma matrices in the spirit of
a Gross-Neveu model \cite{Gross,Neveu(1974)}.
\begin{equation}
     S = im\bar{\Psi}\Psi + \beta\bar{\Psi}\Gamma_\mu\Psi
     \bar{\Psi}\Gamma_\mu\Psi
\end{equation}

This model can be solved by introducing a bosonic variable $\sigma_\mu$ to
remove the 4th degree term

\begin{equation}

    S = im\bar{\Psi}\Psi + 2\beta\bar{\Psi}\Gamma_\mu\Psi \sigma_\mu -
     \beta\sigma_\mu\sigma_\mu + (N/2) ln(2 \pi \beta)
\end{equation}

The fermionic variables can then be integrated giving the determinant of
a matrix whose eigenvalues are easily derived.
\begin{equation}

  Z  = (2 \pi \beta)^{N/2}{\LARGE\int} d\sigma^N {\LARGE|} i m I + 2 \beta
   \Gamma_\mu \sigma_\mu{\LARGE|} exp( -\beta \vec\sigma^2 )\\

  Z  = (2 \pi \beta)^{N/2}{\LARGE\int} d\sigma^N ( 4 \beta^2 \vec\sigma^2 +
            m^2 )^M exp( -\beta \vec\sigma^2 )\\
                  M = 2^{N/2-1}
\end{equation}
which can be reduced to an integral over one variable,
\begin{equation}

   Z  = \beta^{N/2} \Gamma(N/2)^{-1}{\LARGE\int_0^\infty} d\sigma ( 4 \beta^2
       \sigma^2 + m^2 )^M \sigma^{N-1} exp( -\beta \sigma^2 )

\end{equation}

By integrating completely we destroy any possibility of symmetry breaking.
It is necessary to introduce some kind of symmetry breaking term and
rework. Two interesting possibilities are vector terms and matrix terms.

By $O(N)$ invariance a vector term can be rotated to have just one component.
So add a term to the action of the form
\begin{equation}
      S_1 = \bar{\Psi}\Gamma_1\Psi a
\end{equation}
then,
\begin{equation}
       Z  = \beta^{N/2} \Gamma((N-1)/2)^{-1}
       {\LARGE\int_0^\infty} d\sigma
       {\LARGE\int_{-\infty}^\infty} d\sigma_1
       ( 4 \beta^2 \sigma^2 + (2 \beta \sigma_1 + a ) ^2 + m^2 )^M
       \sigma^{N-2} exp[ -\beta (\sigma^2 + \sigma_1^2) ]
\end{equation}
The integrand has two maxima in $\sigma_1$ which dominate the integral.
The asymmetry introduced by the vector term causes a shift from one
maxima to the other and dynamically breaks the symmetry. Taking the
limit $a \rightarrow 0$ indicates that spontaneous breaking of
symmetry can arise.

The result is symmetry breaking from $O(N)$ to $O(N-1)$. Although this is
far from being what we are looking for, a mechanism which selects
one event would be interesting if that event could be identified as the
initial event of the universe!
c.f. \cite{Moffat(1992);Hawking,Laflamme,Lyons(1993)}.

\section{Event Symmetric String Models}

The fact that a large number of degrees of freedom are perhaps required to
produce event symmetry breaking suggests that string theories might
provide answers. The most natural place to start is with
string groups \cite{Kaku(1988)}. These can be formulated without
reference to a
space-time background and can be naturally placed in an event symmetric
setting.

Several models based on discrete string groups will be described but
not in complete detail.

\subsection{Discrete Open Strings}

A basis for a discrete open string algebra is defined by the set of
open ended oriented strings through an event symmetric space of $N$ points.
E.g. a possible basis element might be written,
\begin{equation}

              C = (1,4,3,1,7)

\end {equation}
Note that a string is allowed to cross itself.
The product $A \wedge B$ of two strings in
the algebra is defined by joining them when the end of $A$ matches the
beginning of $B$ reversed and is made to be antisymmetric. Here are some
examples to clarify special cases,
\begin{equation}
        (1,4,3,1,7) \wedge (7,1,3,2,6)  = (1,4,2,6)\\
	(1,4,3,1,7) \wedge (2,3,4,1)  = -(2,3,3,1,7)\\
	(1,4,3,1,7) \wedge (1,6,6)  = 0\\
	(1,4,3,1,7) \wedge (7,5,4,1)  = (1,4,3,1,5,4,1) - (7,5,3,1,7)\\
	(1,4,3,1,7) \wedge (7,1,3,4,1)  = 0\\
	(1,4,3,1,7) \wedge (7,1,3,5,3,4,1)  = (1,4,5,3,4,1) - (7,1,3,5,1,7)\\
\end{equation}
This product satisfies the Jacobi identity,
\begin{equation}

    A \wedge (B \wedge C) + B \wedge (C \wedge A) + C \wedge (A \wedge B) = 0

\end{equation}
The algebra is an infinite dimensional Lie Algebra and it
defines a group by exponentiation. There is also a scalar product
which can be defined. The scalar product is equal to +1 for two base strings if
one is the reverse of the other and zero otherwise. e.g.
\begin{equation}
                    (1,4,3,1,7) \bullet (7,1,3,4,1) = 1\\
		    (1,4,3,1,7) \bullet (3,4,4) = 0
\end{equation}

The reversal of string will be used to define conjugation denoted with
a dagger. Then for any base string C.
\begin{equation}
		    C^\dagger \bullet C = 1
\end{equation}

To define an event symmetric field theory with this discrete string
symmetry take field variables $\Phi$ from the adjoint representation.
In the basis used to define the group the components of $\Phi$ can be
regarded as a family of tensor fields and we can write,
\begin{equation}
   \Phi = \phi() + \phi_\mu (\mu) + \phi_{\mu\nu} (\mu, \nu) + \ldots
\end{equation}

Invariant actions can be constructed using the "vector" and "scalar" products,
\begin{equation}
   S  =  m \Phi \bullet \Phi  +
   \beta (\Phi \wedge \Phi) \bullet (\Phi \wedge \Phi)
\end{equation}

This action presents an immediate problem: It is not positive definite.
Examining the string algebra it is observed that there is a sub-algebra
spanned by the rank two bases,
\begin{equation}
  (\mu, \nu) (\lambda \kappa) = \delta_{\nu\lambda} (\mu, \kappa) -
                                \delta_{\mu\kappa} (\lambda, \nu)
\end{equation}

This is isomorphic to the Lie algebra of the general linear group $GL(N)$
Matrix models based on this group would also have problems over positive
definitiveness of the action. Evidently we must find the string groups which
are extensions of the compact groups such as $SO(N)$ in the same way as this
group is a string extension to $GL(N)$. The algebra will at least have to
be reduced so that,
\begin{equation}
        (\mu, \nu) = -(\nu, \mu)
\end{equation}
This can be extended so that each string is the negative of its conjugate.
That would eliminate the basis elements of rank one altogether and to
keep the algebra closed it is then necessary to eliminate all the odd
rank elements. In this way a consistent algebra is defined with just even rank
tensors and with it
an invariant positive definite action can be constructed as above. This
gives us the first event symmetric string model.

It is important to recognise that the model has an infinite number of
degrees of freedom even for finite $N$. It would be necessary to demonstrate
that it can give a well defined model despite this.

It might be useful to
try and keep the odd rank elements. An algebra can be constructed on
the principle that odd rank base elements are equal to their conjugate
rather than the negative. This does not lead to a positive definite
action.

\subsection{Supersymmetric String Groups}

An attractive feature of the discrete string groups on event symmetric
space-time is that supersymmetric
versions can be defined in a very natural way.

The open string extension of $GL(N)$ can be made supersymmetric in a
straightforward way. The Infinite dimensional Lie algebra is modified
to a graded Lie algebra by identifying the bases of odd rank with odd
elements and those of even rank with even elements. The Lie product of two
odd elements is then made symmetric instead of anti-symmetric.

The same can not be applied to the string extension of the orthogonal group.
The situation is analogous to that of the matrix groups where it is
necessary to go first to the Unitary groups.

The adjoint representation for the open string extension of $U(N)$ is a family
of even rank tensors of complex numbers which become complex conjugate
when the indices are reversed,
\begin{equation}
                 A_{\mu\nu} = A^*_{\nu\mu}\\
                 A_{\mu\nu\lambda\kappa} = A^*_{\kappa\lambda\nu\mu}\\
		 etc.
\end{equation}
Multiplication is a straightforward generalisation of the orthogonal
string group rules and of the adjoint representation of the unitary
groups.

To generalise further to the supersymmetric case odd rank
tensors are introduced, but these must be anticommuting variables.
\begin{equation}
                 \phi_\mu\\
                 \phi_{\mu\nu\lambda}\\
		 etc.
\end{equation}
This supersymmetric generalisation is an analogue of the supersymmetric
generalisation of matrix models already described. A positive definite
action can again be defined on the adjoint representation. This model
may prove to be one of the most important models on Event Symmetric
space-time because of its high symmetry and its obvious relevance to
super-string theories.

The adjoint representation is used by preference since it has a
one-to-one correspondence between field variables and dimensions
of the symmetry group. In principle other representations could be used.

\subsection{Discrete Closed String Groups}

It is also possible to construct Closed String extensions to some
of the matrix groups in which the base elements are cyclically
symmetric. There is no obvious closed string extension to $GL(N)$
but it is easy to extend $SO(N)$. The extension uses only closed
discrete strings of even length with the relations,
\begin{equation}
        (\mu, \nu) = -(\nu, \mu)\\
        (\mu, \nu, \lambda, \kappa) = -(\nu, \lambda, \kappa, \mu)\\
	etc.
\end{equation}
The base elements are multiplied by identifying common sequences
in opposite sense within them.
\begin{equation}
            (1,2,3,4,5,8) \wedge (6,4,3,2,7,9) = (1,6,9,7,5,8)
\end{equation}
This defines an antisymmetric product satisfying the Jacobi Identity.
Supersymmetric extensions also exist for closed strings in which
odd length cycles are included.
\begin{equation}
        (\mu, \nu, \lambda) = (\nu, \lambda, \mu)\\
	etc.
\end{equation}

\subsection{Signature Groups}

Another class of groups closely related to the string groups is
based on sets of discrete events where the order does not matter
accept for a sign factor which changes according to the signature
of permutations,
\begin{equation}
        (\mu, \nu, \lambda) = -(\nu, \mu, \lambda)\\
	etc.
\end{equation}
Multiply by cancelling out any common events with appropriate sign factors.
To get the sign right, permute the events until the common ones are at the
end of the first set and at the start of the second in the opposite
sense. The elements can now be multiplied with the same rule as for the
open string. For this to be consistent with anti-symmetry for even sets
and symmetry for odd sets it is necessary to cancel cases where the
number of common elements is even.
The representations of these groups are families of fully antisymmetric
tensors. The Lie algebras are finite dimensional and the
models based on invariants of these representations are
closely related to the spinor models.

\subsection{Multi-loop String Groups}

Finally there are likely to be multiloop discrete string groups in which more
complex permutation relations are used to represent collections
of loops.

Further classification of these groups and better understanding of
their interrelationships is called for.

It might be hoped that there is a unique model in an adjoint representation
based on these string groups given requirements for a positive definite action
and supersymmetry, This would be consistent with similar conclusions for
continuum string theory \cite{Berkovits(1993)}.

Whether of not interesting spontaneous symmetry breaking mechanisms
can be present in these models is an open question.

\section{Conclusions}

According to the classification of Isham \cite{Isham(1993)} the Event
Symmetric approach to quantum gravity would be a type IV scheme. A new
perspective is proposed from which, it is hoped, continuous
space-time, particle physics and quantum gravity arise. Any such scheme
is necessarily ambitious yet the concept of event symmetric space-time is
both simple and in keeping with previous attempts to quantise gravity.

Symmetric Random Graph models may provide insight into the nature of
space-time,
with dynamic topology and dimensionality. It may be possible by mean
field analysis on symmetric lattices to contrive models which exhibit
changes of dimensionality. Further analysis and numerical simulations
could then improve understanding of such mechanisms.

Of all the models described here the Event Symmetric Superstring models
are the most interesting. Such models have very high symmetry and
may prove to have a rich mathematical structure. There may be
relationships between string theories on Event Symmetric space-time
and string theories on Infinite Dimensional space-time \cite{Juriev(1994)}

There may be a further class of models inspired by the
principle that knot theory is important in quantum gravity
\cite{Witten(1989),Atiyah(1990),Kauffman(1991),Baez(1992)}. The Symmetric
group $S(N)$ could be replaced by the braid group $B(N)$ as the basis
of Event Symmetry. Models based on representations and invariants of
Quantum Groups could be of particular interest.

It is also likely that the concepts of Event Symmetric space-time may
be of value in the study of non-commutative geometries. Indeed such
ideas are already present in recent work \cite{Dimakis,Mu"ller-Hoissen(1994)}.

\end{document}